\documentclass[review]{elsarticle}

\usepackage{lineno,hyperref}

\modulolinenumbers[5]

\usepackage{graphicx}
\usepackage{dcolumn}
\usepackage[geometry]{ifsym}
\usepackage{amsmath}
\usepackage{color}

\begin{document}

\begin{frontmatter}

\title{Two-level systems in evaporated amorphous silicon}

\author[BP]{D.R. Queen\corref{mycorrespondingauthor}\footnote{{\it Present Address:} Northrop-Grumman Electronic Systems, Lithicum, MD 21240, USA}}
\cortext[mycorrespondingauthor]{Corresponding author}
\ead{daniel.queen@ngc.com}

\author[NRL]{X. Liu}

\author[BM]{J. Karel}

\author[BP]{H.C. Jacks}

\author[NRL]{T.H. Metcalf}

\author[BP,BM]{F. Hellman}

\address[BP]{Department of Physics, University of California, Berkeley, Berkeley, CA 94720, USA}
\address[NRL]{Naval Research Laboratory, Washington D.C., 20375, USA}
\address[BM]{Department of Materials Science and Engineering, University of California, Berkeley, Berkeley, CA 94720, USA}


\begin{abstract}

In $e$-beam evaporated amorphous silicon ($a$-Si), the densities of two-level systems (TLS), $n_{0}$ and $\overline{P}$, determined from specific heat $C$ and internal friction $Q^{-1}$ measurements, respectively, have been shown to vary by over three orders of magnitude. Here we show that $n_{0}$ and $\overline{P}$ are proportional to each other with a constant of proportionality that is consistent with the measurement time dependence proposed by Black and Halperin and does not require the introduction of additional anomalous TLS. However, $n_{0}$ and $\overline{P}$ depend strongly on the atomic density of the film ($n_{\rm Si}$) which depends on both film thickness and growth temperature suggesting that the $a$-Si structure is heterogeneous with nanovoids or other lower density regions forming in a dense amorphous network. A review of literature data shows that this atomic density dependence is not unique to $a$-Si. These findings suggest that TLS are not intrinsic to an amorphous network but require a heterogeneous structure to form.

\end{abstract}

\end{frontmatter}

\section{Introduction}

At low temperatures, the thermal, acoustic, and dielectric properties of amorphous materials are dominated by low energy excitations that are not normally found in crystalline materials.~\cite{Zeller1971} It was originally believed that these excitations were intrinsic to the amorphous state since they occurred with roughly the same density in all amorphous solids and were independent of the chemical constituents of the materials.~\cite{Stephens1973,Pohl2002} However, recent measurements have shown that TLS can be suppressed in certain thin film materials suggesting that the vapor deposition process leads to a fundamentally different energy landscape than quenching from a liquid.~\cite{Liu1997,Liu1998,Swallen2007,Ashtekar2011,Queen2013,Liu2014,Castenada2014} The two-level systems (TLS) model successfully describes many low temperature phenomena, such as the linear temperature dependence of the specific heat and the internal friction plateau. The TLS model is an incomplete description of the amorphous state though as it does not describe other ubiquitous low temperature phenomena, such as the excess $T^{3}$ specific heat or the thermal conductivity plateau.~\cite{Zeller1971,Phillips1972,AHV1972} 

Systematically testing the TLS model in amorphous solids has proven difficult as the physical origin of the TLS in these materials is unknown. Understanding the origin of these excitations and how to controllably remove them has gained new urgency as decoherence caused by TLS is a major roadblock for quantum devices.~\cite{Martinis2005,Gao2007} Recently we have shown that the TLS in $a$-Si can be removed by increasing the atomic density of the film, which depends on both the film thickness $t$ and growth temperature $T_{S}$, and that the TLS are linked to the excess $T^3$ heat capacity.~\cite{Queen2013,Liu2014} In this article, we use the tunability of the TLS in $a$-Si to show that the TLS are described by the standard TLS model and that anomalous TLS, those that contribute to $C$ and not $Q^{-1}$, are not required to explain our results.  

The TLS model assumes that neighboring minima in the potential energy landscape of an amorphous solid can be treated as double-well potentials where the wave function overlap between wells creates a finite tunneling probability which splits the ground state energy creating the TLS.~\cite{Phillips1972,AHV1972} The TLS must be broadly distributed in the energy landscape to describe the experimental results. Physically, the TLS are thought to correspond to single atoms or groups of atoms with energetically similar configurations that are separated by energy barriers on the order of $100$K with tunnel splittings $<1$K.~\cite{Hunklinger1974} It has been suggested that the amorphous structure must be open and have low coordination for the TLS to form.~\cite{Phillips1972} Structural rigidity is known to play a key role in the glass forming ability and elastic properties of amorphous solids~\cite{Boolchand2000} and the rigid four fold coordination in tetrahedrally bonded materials, such as $a$-Si and $a$-Ge, is generally thought to prevent the formation of TLS.~\cite{Phillips1972} There are conflicting results in the literature on whether TLS occur in $a$-Si and $a$-Ge and it has been debated whether the TLS in these systems are the same as those found in other glasses, such as $a$-SiO$_{2}$.~\cite{Haumeder1980,Lohneysen1982,Mertig1984,Graebner1984a,Graebner1984,VandenBerg1985,Zink2006a} The TLS density in tetrahedrally bonded materials depends strongly on the preparation technique suggesting that the TLS are due to some microstructural detail of the material.~\cite{Liu1998,Liu2002} For example, it was previously thought that hydrogen played a key role in removing TLS from $a$-Si~\cite{Liu1997} but our recent results show that the reduction in TLS in the hydrogenated material was likely the result of increasing $T_{S}$.~\cite{Liu2014} 

The presence of low energy excitations can be seen experimentally in the low temperature specific heat $C$ which, for an amorphous dielectric, has the form,~\cite{Stephens1973}
\begin{equation}
\label{eq:TLS_Model}
C=c_{1}T+c_{3}T^{3}
\end{equation}
where $c_{1}$ is the linear specific heat due to the low energy excitations. In the TLS model, $c_{1}$ for $a$-Si, in units of J~mol$^{-1}$~K$^{-2}$, is expressed as
\begin{equation}
c_{1}=\frac{\pi^2}{6} k^{2}_{B} n_{0}\frac{N_{A}}{n_{\rm Si}},
\label{eq:TLS_C_term}
\end{equation}
where $k_{B}$ is Boltzmann's constant, $n_{0}$ is the density of TLS, $N_{A}$ is Avogadro's number, and $n_{\rm {Si}}$ is the Si number density.~\cite{Phillips1981a}  $n_{0}\approx 10^{45}$J$^{-1}$m$^{-3}$ for most glasses.~\cite{Stephens1973} 
\begin{equation}
c_{3}=c_{D}+c_{ex}
\label{eq:cubic_term}
\end{equation}
is larger than the specific heat due to phonons $c_{D}$. $c_{ex}\approx c_{D}$ and is thought to be caused by excess, non-propagating vibrational modes. 

At low $T$, the interaction between the TLS and elastic/thermal waves leads to loss in acoustic measurements~\cite{Hunklinger1976} and the $T^{2}$ temperature dependence of the thermal conductivity~\cite{Zeller1971,Stephens1973}. The interaction between acoustic waves and TLS leads to dissipation, or internal friction $Q^{-1}$, and is observed as a temperature independent plateau in at $T\approx0.1-10$K in $Q^{-1}$. In the TLS model, the plateau $Q^{-1}_{0}$ is expressed as
\begin{equation}
Q^{-1}_{0}=\frac{\pi}{2}\frac{\overline{P}\gamma^{2}_{i}}{\rho v^{2}_{i}},
\label{eq:IF_TLS_Model}
\end{equation}
where $i$ indicates the polarization of the wave (longitudinal or transverse), $\overline{P}$ is the spectral density of TLS, $\gamma_{i}\sim 0.1-1$~eV is the coupling energy between TLS and acoustic waves~\cite{Berret1988}, $\rho$ is the mass density, and $v_{i}$ is the sound velocity. $Q^{-1}_{0}\approx 10^{-4}$ ($\overline{P}\approx10^{44}$J$^{-1}$m$^{-3}$) and varies little for bulk quenched glasses of differing chemical composition for both wave polarizations.~\cite{Pohl2002} The TLS density measured by $Q^{-1}$, acoustic attenuation, or thermal conductivity is often described as ``universal'' due to the insensitivity of the measurement results to chemical composition and excitation frequency~\cite{Pohl2002} and is often regarded as a measure of the intrinsic TLS described by the TLS model.~\cite{Berret1988}

Comparison of the TLS densities measured by $C$ and $Q^{-1}$ show that $n_{0} \approx 20 \overline{P}$ for the materials where both have been measured.~\cite{Zimmermann1981,Meissner1981,Loponen1982,Zeller1971,Stephens1973,Berret1988,Grace1989} Measurements such as $Q^{-1}$ are performed in the frequency domain with an excitation frequency $\omega$. These measurements yield the spectral TLS density as only those TLS with relaxation time $\tau$, satisfying $\omega\tau \sim 1$, are probed. Specific heat measurements on the other hand measure all TLS that can equilibrate with the phonon bath on the time scale of the measurement. Black and Halperin considered how a distribution of relaxation times in the TLS model would affect $C$ measurements~\cite{Black1977,Black1978} and found that
\begin{equation}
n_{0}=\frac{1}{2}\overline{P}\textrm{ln}\left (\frac{4\tau}{\tau_{\textrm{min}}}\right ),
\label{eq:TLS_time_relation}
\end{equation}
where $\tau$ is the measurement time and $\tau_{\rm min}$ is the minimum TLS relaxation time which is estimated to be $\approx 10^{-9}$~sec from comparison of $n_{0}$ and $\overline{P}$ for $a$-SiO$_{2}$.~\cite{Black1977} It is assumed that $\tau_{\rm min}$ varies little between amorphous materials. The predicted logarithmic time dependence was found in $C$ measurements where $\tau<100$~$\mu$sec.~\cite{Loponen1982} However for longer $\tau$, $C$ increased faster than predicted by Eq.~\ref{eq:TLS_time_relation} which suggested either a non-uniform spectral distribution of TLS~\cite{Loponen1982} or that additional anomalous TLS contribute to $C$ at longer $\tau$.~\cite{Black1978,Loponen1982}

In this paper, we compare the TLS densities $n_{0}$ and $\overline{P}$ from $C$ and $Q^{-1}$ measurements, respectively, of $e$-beam evaporated $a$-Si films and show that the low energy excitations are described well by the TLS model over three orders of magnitude in TLS density. Both $C$ and $Q^{-1}$ at low $T$ are typical of amorphous materials containing TLS: $C$ is linear in temperature and $Q^{-1}$ has a temperature independent plateau. Both measures of the TLS density are found to depend strongly on the atomic density of the film $n_{\textrm{Si}}$ which varies with deposition temperature $T_{S}$ and film thickness. The highest density films have TLS densities near or below the detection level of either technique, while lower density films have a significant TLS density that is similar to other amorphous solids. The agreement between $n_{0}$ and $\overline{P}$ shows that additional, anomalous TLS are not required to explain our results. We suggest that $a$-Si has a heterogeneous structure consisting of voids, or some other low density structure, surrounded by a dense backbone network. Finally, we compare our $a$-Si results to literature data on other materials where the TLS density was found to vary and atomic density was available. We observe a qualitatively similar density dependence in these systems which suggests that TLS in the low coordination bulk glasses are not an intrinsic result of disorder but also depend on the nano-scale structure of the materials. 

\section{Experimental Procedure}

$A$-Si thin films were prepared by \textit{e}-beam evaporation at a base pressure of $\sim1\times10^{-8}$~Torr and a growth rate of $0.05-0.1$nm/s. Growth temperature $T_{S}$ was varied from $45^{\circ}-400^{\circ}$C. Films were grown separately on membrane-based nanocalorimeters and single crystal double paddle oscillators (DPO). The film thickness $t$ was varied for the $C$ measurements while the $Q^{-1}$ films were all nominally 300nm thick to ensure an adequate measurement signal. Thicknesses were measured on films grown on neighboring substrates using a KLA-Tencor Alpha-Step IQ profilometer with an error of $1\%-4\%$ depending on the film thickness. The uncertainty in $t$ is the dominant source of error except in some samples where $C$ below 10K was less than $10$\% of the total measured heat capacity; in that case random error in the measurement dominates. The average film densities were determined from Rutherford backscattering (RBS). All of the films were found to have a thin surface oxide ($1-2$nm). The lower density films grown at $45^{\circ}$C films had $\approx 4-5$ at.~\% oxygen below the surface while the higher $T_{S}$ films had $\leq 1$~at.~\% oxygen with concentration profiles consistent with post-deposition diffusion into the film. 

\begin{figure}%
\includegraphics{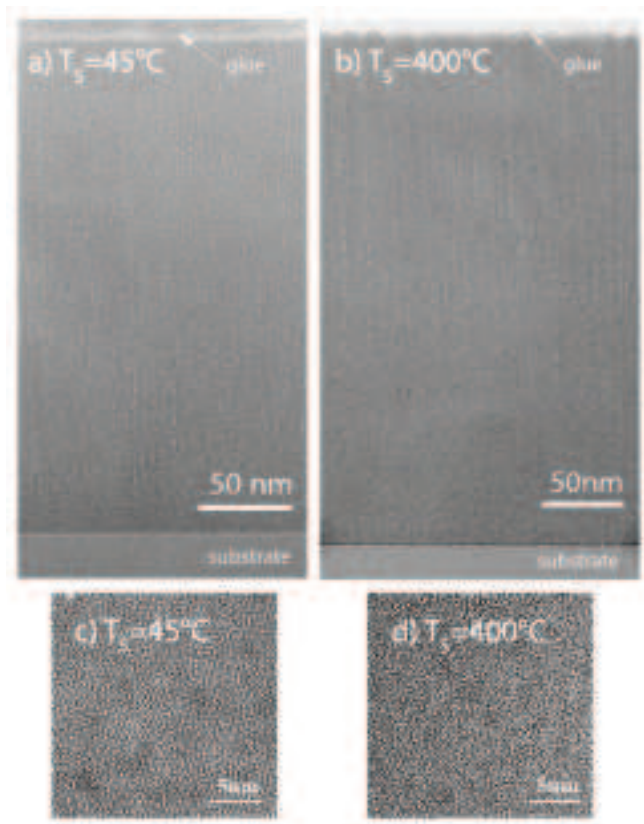}%
\caption{Cross sectional TEM a) T$_{S}$=45$^{\circ}$C ($t=278$nm) and b) 400$^{\circ}$C ($t=310$nm). High resolution TEM are shown in c) and d) for the same films, respectively.}%
\label{fig:TEM}%
\end{figure}

Transmission electron micrographs (TEM) were taken in cross sections on films grown at $45^{\circ}$C and $400^{\circ}$C and are shown in Fig.~\ref{fig:TEM}. The low magnification TEM shows a columnar growth structure, which is commonly seen in evaporated films, with the $T_{S}=400^{\circ}$C film having larger diameter columns ($8\pm5$~nm) than the $45^{\circ}$C film ($4\pm3$nm). The column diameters appear constant throughout the thickness of the films. The high resolution TEM shows that the films are fully amorphous with no lattice fringes or diffraction peaks. Similarly, X-ray diffraction showed no peaks. Dangling bond densities $n_{\rm ESR}$ were determined from electron spin resonance(ESR) measurements where the gyromagnetic ratio $g=2.0055$ as is typical for isolated, neutral dangling bonds in $a$-Si.~\cite{Thomas1978} Raman scattering measurements were performed using the 514.5 nm line of an Ar ion laser.\cite{Queen2013}

Heat capacity measurements were made from $2-300$K using a microfabricated nanocalorimeter. Details of the measurement technique are provided elsewhere.~\cite{Queen2009,Denlinger1994,Revaz2005} For these measurements a 20nm~\textit{a}-AlO$_{x}$ diffusion barrier was sputtered onto the $a$-Si sample before deposition of the Cu film which is used to ensure the sample is isothermal during the measurement. The calorimeter with only the $a$-AlO$_x$ and Cu films was measured separately for subtraction of the background heat capacity.

$Q^{-1}$ measurements were made from $0.3-300$K using the anti-symmetric torsional mode of the DPO at 5500Hz.~\cite{White1995b} The shift in the resonant frequency of DPO after the deposition of the film gives the shear modulus $G=\rho v_{t}^2$ where $v_{t}$ is the transverse sound velocity. The longitudinal sound velocity $v_{l}$ was measured using an ultrasonic pump/probe technique.~\cite{Lee2005}

\section{Results}

\begin{figure}[h]%
\includegraphics[scale=0.8]{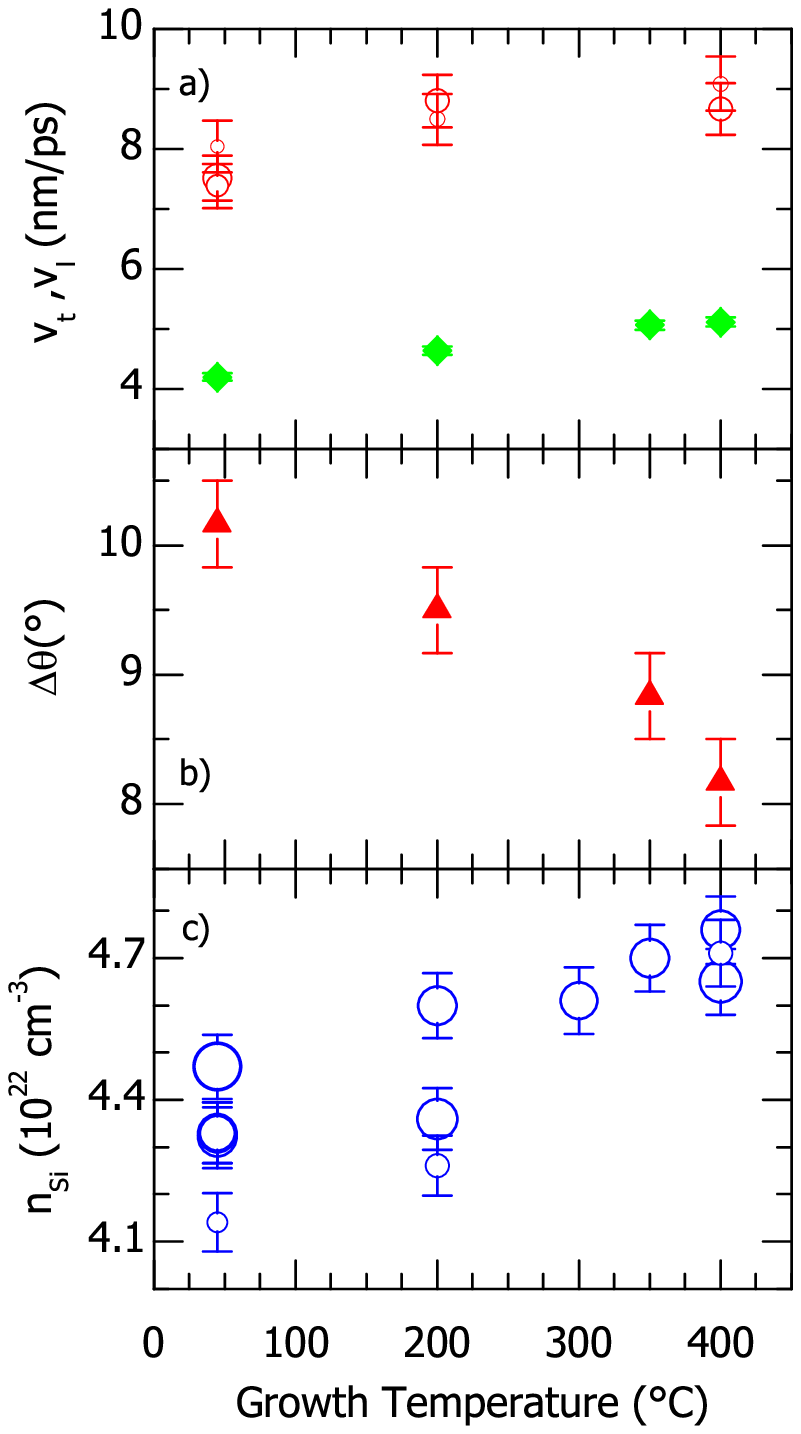}%
\caption{Growth temperature dependence of the a) longitudinal (circles) and transverse (diamonds) sound velocity, b) bond angle disorder $\Delta \theta$, and c) silicon density. Symbol size represents the relative thickness of each film which varies between 100nm and 400nm. The $v_{t}$ and $\Delta\theta$ samples are $\approx 300$nm thick.}%
\label{fig:MP}%
\end{figure}

Figure~\ref{fig:MP} shows $v_{l}$, $v_{t}$ determined from $G$, bond angle disorder $\Delta \theta$, and $n_{\rm Si}$ all as functions of $T_{S}$. The data are given in Table~\ref{tab:table1} and Table~\ref{tab:table2}. The symbol size represents the relative thickness of each film which ranges between 100nm and 400nm. Both $v_{l}$ and $v_{t}$ increase with $T_{S}$ indicating that the bonds in the amorphous network become stiffer as $T_{S}$ increases. $\Delta \theta$, which is determined from the width of the transverse optic-like peak in the Raman spectrum~\cite{Queen2013Sup}, decreases with increasing $T_{S}$ indicating that the bonds are more ordered in the higher $T_{S}$ films. $n_{\rm Si}$ generally increases with $T_{S}$ and $t$. These results suggest that increasing $T_{S}$ improves the structural order in the amorphous network.

\begin{figure}[t]
	\centering
	\includegraphics{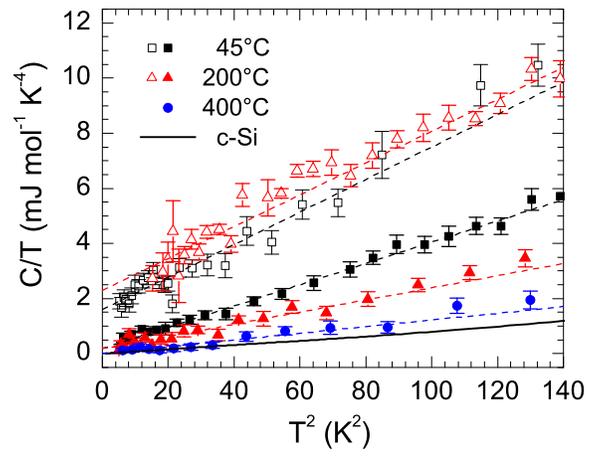}
	\caption{Specific heat of $a$-Si plotted as $C/T$ versus $T^{2}$. Dashed lines are fits to the data using Eq.~\ref{eq:TLS_Model}. Open symbols are the thinner films having lower $n_{\textrm {Si}}$. Crystalline silicon (solid line) is shown for comparison.}
	\label{fig:CoverT}
\end{figure}

\begin{table}[h]

\caption{\label{tab:table1}Summary of data: growth temperature $T_{S}$, sample thickness $t$, silicon number density $n_{Si}$ (for crystalline Si $n_{Si}=5.00\times10^{22}$cm$^{-3}$), dangling bond density $n_{ESR}$, and longitudinal sound velocity $v_{l}$. $\theta_{D}$ is the Debye temperature calculated from $v_{l}$ and $v_{t}$ with $c_{D}$ the corresponding $T^{3}$ specific heat. $c_{1}$ and $c_{3}$ are from fits to the $C$ data at low $T$ using Eq.~\ref{eq:TLS_Model}. $n_{0}$ is the density of TLS from $c_{1}$. $c_{ex}=c_{3}-c_{D}$ is the excess $T^{3}$ specific heat.}
\resizebox{12.5cm}{!}{
\begin{tabular}{ccccccccccc}
$T_{S}$ &	$t$ &	$n_{Si}$ &	$n_{ESR}$ &	$v_{l}$ &	$\theta_{D}$ &	$c_{D}$ &	$c_{1}$ &	$c_{3}$ &	$n_{0}$ &	$c_{ex}$\\
\hline
&	&	$\times 10^{22}$ &	$\times 10^{18}$ &	&	&	$\times 10^{-5}$ &	$\times 10^{-4}$ &	$\times 10^{-5}$ &	$\times 10^{46}$ &	$\times 10^{-5}$ \\
$^{\circ}$C &	nm &	cm$^{-3}$ &	cm$^{-3}$ &	nm ps$^{-1}$ &	K &	J mol$^{-1}$~K$^{-4}$ &	J mol$^{-1}$~K$^{-2}$ &	J mol$^{-1}$~K$^{-4}$ &	J$^{-1}$~m$^{-3}$ &	J mol$^{-1}$~K$^{-4}$ \\
\hline
45 & 	112 & 	4.14 &	 6.6 &	8.04 & 	476 & 	1.8 &  	16 & 	5.9 &	36 & 	4.1\\
45 &	278 &	 4.33 &	 6.0 & 	7.38 & 	479 & 	1.8 & 	1.9 &	 3.9 &	 4.6 & 	2.1\\
200 & 	153 & 	4.26 &	 6.7 &	8.49 & 	550 & 	1.2 & 	8.0 &	 5.4 & 	18 & 	4.2\\
200 & 	319 &	 4.36 &	 5.8 & 	8.80 & 	560 & 	1.1 & 	1.8 & 	2.2 &	 4.3 & 	1.1\\
400 & 	310 & 	4.71 &	 5.1 & 	8.66 &	611 & 	0.9 & 	0.1 & 	1.2 &	 0.2 & 	0.4\\
\hline

\end{tabular}
}
\end{table}

Figure~\ref{fig:CoverT} shows $C$ at low $T$ for several $a$-Si films~\cite{Queen2013} with different thicknesses and growth temperatures plotted as $C/T$ vs $T^2$ along with fits (dashed lines) to Eq.~\ref{eq:TLS_Model} which appear as straight lines on this plot. Crystalline silicon is shown for comparison.~\cite{Touloukian}  The intercept of the fit corresponds to $n_{0}$ and is due to TLS while the slope corresponds to $c_{3}$ and is due to both phonons ($c_{D}$) and non-propagating vibrational modes ($c_{ex}$). As noted above, the sound velocities, and thus $c_{D}$, depend only on $T_{S}$ (Fig.~\ref{fig:MP}a). $C$ however is not monotonic in $T_{S}$ but also varies with $t$ as was found with $n_{\rm Si}$. We note that we do not expect changes in $C$ due to a dimensionality crossover (i.e. 3D to 2D) as $t$ decreases. The dominant phonon approximation can be used to calculate the frequency of the dominant heat carrying phonons $\nu_{dom}=90{\rm GHz/K} T$. At $2$K, the phonon wavelength $\lambda_{dom}=v/\nu_{dom}$ is $20 - 50$ nm depending on the polarization and sound velocity and is less than $t>100$nm. At high $T$, the data all converge toward the Dulong-Petit limit (25 J mol$^{-1}$K$^{-1}$) indicating that the differences in $C$ are not due to an error in $n_{\rm Si}$.

\begin{table}
\caption{\label{tab:table2}Summary of data: $T_{S}$ is the growth temperature, $n_{\rm Si}$ is the silicon number density, $G$ is the shear modulus, $v_{t}$ is the transverse sound velocity, and $Q^{-1}_{0}$ is the magnitude of the internal friction. $\overline{P}\gamma^{2}$ is calculated from Eq.~\ref{eq:IF_TLS_Model}. $\overline{P}$ is calculated assuming $\gamma=0.3$ev. Film thickness is $300$nm for all samples.}
\begin{tabular}{ccccccc}
$T_{S}$	&	$n_{\rm Si}$	&	$G$	&	$v_{t}$	&	$Q_{0}^{-1}$	&	$\overline{P}\gamma^2$	&	$\overline{P}$	\\
\hline
	&	$\times 10^{22}$	&		&		&	$\times 10^{-5}$	&	$\times 10^{6}$	&	$\times 10^{44}$	\\
	&	cm$^{-3}$	&	GPa	&	nm ps$^{-1}$	&		&	J m$^{-3}$	&	J$^{-1}$m$^{-3}$	\\
	\hline
45	&	4.32	&	33.2	&	4.06	&	18.74	&	6.22	&	27.0	\\
45	&	4.33	&	35.1	&	4.17	&	17.11	&	6.01	&	26.1	\\
200	&	4.60	&	46.1	&	4.64	&	5.68	&	2.62	&	11.4	\\
300	&	4.61	&	43.3	&	4.49	&	4.90	&	2.12	&	9.2	\\
350	&	4.70	&	56.2	&	5.06	&	0.55	&	0.31	&	1.4	\\
400	&	4.65	&	54.7	&	5.02	&	0.30	&	0.17	&	0.7	\\
400	&	4.76	&	58.0	&	5.11	&	0.20	&	0.12	&	0.5	\\
\hline

\end{tabular}
\end{table}

\begin{figure}[t]
	\centering
	\includegraphics[scale = 0.8]{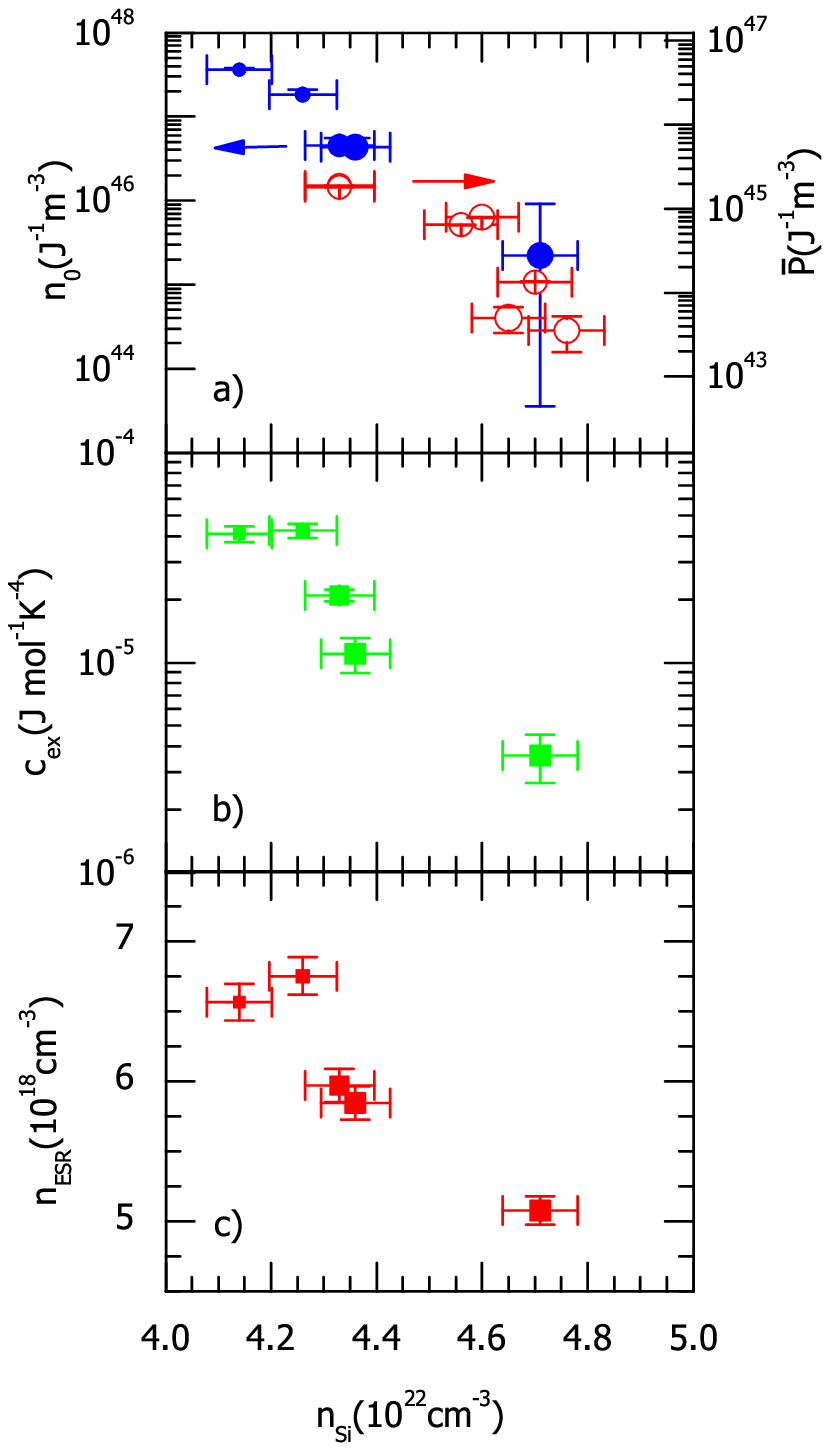}
	\caption{a) Amorphous silicon TLS densities $n_{0}$ (solid circles), determined from $C$, and $\overline{P}$ (open circles), determined from $Q^{-1}$, as a function of $n_{\textrm {Si}}$. The $\overline{P}$ and $n_{0}$ axes are scaled such that $n_{0}=8\overline{P}$ by using Eq.~\ref{eq:TLS_time_relation}. b) $c_{ex}$ due to non-propagating modes. c) Dangling bond density $n_{\rm ESR}$. Symbol size represents the relative thickness of each film which varies between 100nm and 400nm.}
	\label{fig:parameters}
\end{figure}

Figure~\ref{fig:parameters}a compares $n_{0}$ from the fit of $C$ to Eq.~\ref{eq:TLS_Model} to $\overline{P}$ determined from $Q^{-1}_{0}$ using Eq.~\ref{eq:IF_TLS_Model}~\cite{Queen2013,Liu2014}. The data are shown as a function of $n_{\rm Si}$. The TLS acoustic coupling energy $\gamma$ has not been measured for $a$-Si so we use $\gamma=0.36$eV which has been measured for $a$-Ge and is expected to be within a factor of 2 of the real value.~\cite{Duquesne1983} Figure~\ref{fig:parameters}a shows that $n_{0}$ and $\overline{P}$ both depend strongly on $n_{\rm {Si}}$. The axes of Fig.~\ref{fig:parameters} are scaled such that $n_{0}/\overline{P}=8$ as is predicted from Eq.~\ref{eq:TLS_time_relation} when using the specific heat measurement time $\tau \approx 10^{-3}$~sec, which is set by the low $T$ relaxation time of the calorimeter, and assuming $\tau_{\rm min}=10^{-9}$~sec. The data are consistent with the time dependence of Eq.~\ref{eq:TLS_time_relation} within measurement error and the uncertainty in the choice of $\gamma$ which likely varies with sound velocity (and thus $T_{S}$ but not $n_{\rm Si}$)~\cite{Berret1988}. More importantly, the dependence of both $n_{0}$ and $\overline{P}$ on $n_{\rm Si}$ suggests that the states observed in $C$ are the same as those that cause loss in $Q^{-1}$ and should not be considered anomalous. 

Figure~\ref{fig:parameters} also shows the $n_{\rm Si}$ dependence of $c_{ex}$ (Fig.~\ref{fig:parameters}b) and dangling bond density $n_{\rm ESR}$ (Fig.~\ref{fig:parameters}b). As with $n_{0}$ and $\overline{P}$, $c_{ex}$ is higher in the low density films suggesting that the same structures are responsible for both the TLS and the non-propagating modes. The dependence of $C$ on $n_{\rm Si}$, despite the dependence of sound velocity on $T_{S}$, is explained by considering phonons and low energy excitations separately. The TLS and non-propagating modes are both associated with the low density regions, while the phonon term $c_{D}$ depends entirely on the degree of order in the amorphous network. From these results we suggest that $a$-Si has a heterogeneous structure consisting of distinct voids or lower density regions surrounded by a dense backbone network. Increasing $T_{S}$ increases the structural order in the backbone network which we suggest carries the propagating sound waves measured by $v_{l}$ and $v_{t}$. This structural model is supported by the density of dangling silicon bonds $n_{\rm ESR}$ which are a proxy for voids in $a$-Si~\cite{Brodsky1972} and scale linearly with $n_{\rm Si}$ (Fig.~\ref{fig:parameters}c).  

\begin{figure}[t]
	\centering
	\includegraphics{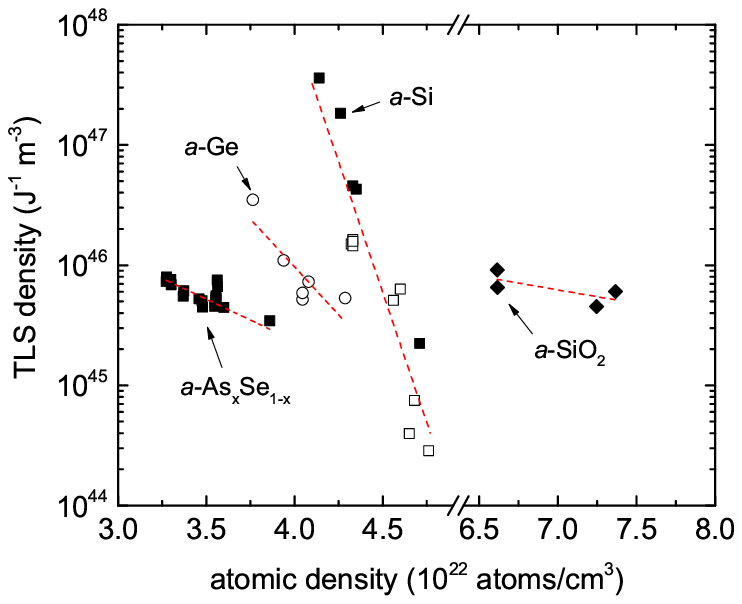}
	\caption{Comparison of TLS density versus atomic number density for several amorphous materials. Filled symbols are $n_{0}$ determined from $C$ measurements and open symbols are $\overline{P}$ from either $Q^{-1}$ ($a$-Si) or thermal conductivity ($a$-Ge). The $\overline{P}$ data for $a$-Si and $a$-Ge are multiplied by a factor of $8$ to put them on the same scale as $n_{0}$ for $a$-Si. Lines are a guide to the eye.}
	\label{fig:TLS_compare}
\end{figure}

The dependence of TLS on mass density has been noted previously but not systematically studied.~\cite{Phillips1981a,Graebner1984,Grace1989,Brand1991} Here we compile literature results for several material systems where either $n_{0}$ or $\overline{P}$ was found to vary and density values were available. Figure~\ref{fig:TLS_compare} compares $n_{0}$ and $\overline{P}$ for $a$-Si from this work to $\overline{P}$ for evaporated $a$-Ge determined from thermal conductivity measurements~\cite{Graebner1984} along with $n_{0}$ from $C$ measurements of $a$-As$_{x}$Se$_{1-x}$~\cite{Liu1993} and $a$-SiO$_{2}$~\cite{Liu1995} all as a function of atomic density. The horizontal axis is broken for clarity and the dashed lines for each material are guides to the eye. The $\overline{P}$ data for $a$-Si and $a$-Ge are multiplied by a factor of $8$ to put them on the same scale as $n_{0}$ for $a$-Si. The measurement times for the $a$-As$_{x}$Se$_{1-x}$ and $a$-SiO$_{2}$ samples are unknown but are likely on the order of $1$~sec. The difference in the $C$ measurement time scales between the bulk and thin film measurements is expected, from Eq.~\ref{eq:TLS_time_relation}, to result in only a factor of $2$ difference when comparing $n_{0}$ values. The atomic density of $a$-Ge and $a$-Si depends on film deposition conditions.  The density of $a$-SiO$_{2}$ (Suprasil I and Suprasil W) was increased by irreversibly compressing the materials at high temperature and pressure.~\cite{Liu1995} The density of $a$-As$_{x}$Se$_{1-x}$ varies with composition $x$~\cite{Feltz1983,Brand1991,Liu1993} and was taken from a separate study cited by the authors. A universal TLS density is not observed in Fig.~\ref{fig:TLS_compare}, but all materials show the same systematic dependence: the TLS density decreases with increasing atomic density. Understanding the origin of the density changes in these materials may elucidate the structures responsible for the TLS.

\section{Discussion}

The microstructure of $a$-Si is known to depend strongly on the film deposition conditions and for our evaporated films we find that $n_{\rm Si}$ changes with both $T_{S}$ and film thickness. We can make several statements about the low density regions without precise knowledge of their local microstructure. First, the inter-columnar regions visible in the TEM~\cite{Queen2013Sup} are likely to have lower density than the columns but it is unlikely that the TLS are solely due to states at the column boundaries as the change in column diameters and thus their surface area is small compared to the orders of magnitude change in the TLS density. We can also rule out a density gradient in the film where $n_{Si}$ increases with $t$ since the $total$ heat capacity at low $T$ of thinner films (in J/K, not normalized by film volume) is $larger$ than that of thicker films. Similarly, we can also rule out oxygen as the TLS as neither $n_{0}$, $\overline{P}$, nor $c_{ex}$ depend on the oxygen content. Figure~\ref{fig:parameters}c shows that $n_{\rm ESR}$ is higher in the low density films with a very similar dependence on $n_{\rm Si}$ as $n_{0}$ and $c_{ex}$. However, $n_{\rm ESR}$ changes by a factor of 2 while $n_{0}$ and $c_{ex}$ vary by a factor of 1000. (Note the linear versus log scales.) In addition, both $C$ and $Q^{-1}$ are insensitive to magnetic fields~\cite{Metcalf2000,QueenThesis} whereas excitations due to dangling bonds/electronic structure would depend strongly on field.~\cite{VandenBerg1985} Thus the dangling bonds cannot be causing the TLS and non-propagating modes but are likely related to the same underlying structure.

A dependence of density on thickness has previously been reported for $a$-Si where it was suggested that a network of interconnected voids occurs in the lower density films.\cite{Bean1979, Foti1980} This result is consistent with the heterogeneous $a$-Si structure that we propose but the origin of the thickness dependence is not understood. It is possible that the material becomes denser in thicker films to relieve stress built up during the deposition process similar to how crystalline films relieve stress by nucleating defects at the film-substrate interface; this interpretation however requires further study.

Bonding constraints have long been thought to play a role in the formation of TLS.  Under-coordinated structures, like the floppy Si--O--Si bridges in $a$-SiO$_{2}$, should be amenable to TLS while rigid structures, like the four-fold coordinated network of $a$-Si, should not. TLS are often thought to result from floppy bonds in a disordered network. Our data show that the TLS can form in the $a$-Si network even though the bonding is over-constrained. Thus if floppy modes are important it may be that a local reduction in rigidity of the structure matters more than the average coordination.~\cite{Alexander1998} Surfaces and voids offer natural locations for such structures to occur. 

In our heterogeneous structural model, the TLS should be localized in the low density regions and not the rigid matrix. Neutron scattering measurements~\cite{Kamitakahara1987} and molecular dynamics simulations~\cite{Nakhmanson2000} both suggest that nano-scale voids may play a role in the occurrence of additional low energy excitations in $a$-Si. These excitations are absent in simulations of full density networks.\cite{Feldman1993,Feldman1999} A heterogeneous structural model for amorphous solids has been proposed which consists of elastically soft regions embedded in a rigid matrix where internal and external stresses play a role in the resulting amorphous structure.\cite{Alexander1998} Our experimental results are in good agreement with the predictions of the model. Namely, the soft regions have lower density than the rigid matrix. Thus macroscopic properties, such as the sound velocity, are due to the rigid matrix while the TLS and excess vibrational excitations are due to the soft regions. Techniques, such as fluctuation electron microscopy, that probe the structure of $a$-Si at $1-2$~nm length scale, may clarify whether this picture of the amorphous state is correct.\cite{Treacy2012,Roorda2012}

The data in Fig.~\ref{fig:TLS_compare} show that the atomic density dependence of TLS found in $a$-Si applies to other glasses as well, suggesting that the structure of these glasses may also be heterogeneous. It may thus be possible to prepare an ``ideal'' glass without TLS, even in non-tetrahedrally bonded traditional glass materials such as $a$-SiO$_{2}$, by eliminating the low density regions. Understanding where these regions are located and how they form is crucial to controlling the TLS density in amorphous solids. No TLS-free amorphous solids have been prepared by rapidly quenching from a liquid to the glassy state. However, the universal TLS density found in most amorphous solids may thus result from heterogeneities and low density regions produced by this quenching process, as suggested by Lubchenko and Wolynes (LW).~\cite{Lubchenko2001,Lubchenko2007} LW relate quench rate, fictive temperature $T_{f}$, atomic density and TLS density in quenched glasses, finding that TLS density goes as $(T_{f}-T_{k})^{2}/T_{f}$, where $T_{k}$ is the Kauzmann temperature. Broadly speaking, in their model, the slower the quench, the more dense the glass, the lower the $T_{f}$, and hence the lower the TLS density.  Although the landscapes of quenched glasses are generally considered to be quite different from those of vapor deposited glasses, this relationship between atomic density and TLS density suggests that they may be more similar than currently believed. Vapor deposition of thin films may provide a method of preparing more ideal structures as surface mobility during deposition has been shown to be crucial for forming dense polymer and tetrahedrally-bonded glasses that are quenched deeply in the amorphous energy landscape.~\cite{Queen2013,Swallen2007,Castenada2014} 

\section{Conclusions}

In summary, using both specific heat and internal friction measurements, we have confirmed that the TLS found in evaporated $a$-Si are described by the TLS model and do not require the introduction of anomalous TLS. The TLS have the same structural origin as the non-propagating modes and are likely due to nano-scale heterogeneity. These regions are distinct from the backbone amorphous network. We find a similar density dependence in other amorphous solids suggesting that these results apply more broadly to other glasses. Theoretical results on liquid quenched glasses show a similar dependence and suggest that the landscapes for vapor deposited and liquid quenched amorphous solids may be more similar than generally recognized, with the differences arising from the more universal nature of liquid quenching through a glass transition, compared to the wide range of factors that influence vapor deposition growth, which we suggest can produce samples that lie either deeper or higher in the energy landscape than liquid quenching processes. A detailed understanding of the physical origin of TLS will be required to remove them from systems that are affected by TLS losses, such as superconducting devices and solid state quantum bits, detectors, and amplifiers.  

We thank V. Lubchenko for useful discussions, K.M. Yu and R. Culbertson for assistance with RBS, D.J. Smith for TEM, J.W. Ager III for assistance with Raman, and D. Bobela for ESR. This work supported by NSF DMR-0907724, film growth and nanocalorimeter development supported by the U.S. Department of Energy DE-AC02-05CH11231, and internal friction measurements supported by the Office of Naval Research.

\section*{References}

\end{document}